\documentclass[11pt]{article}
\usepackage{amsmath,amssymb}
\usepackage[margin=1in]{geometry}

\title{Quantum mechanics provides the physical basis of teleological evolutions}
\author{Giuseppe Castagnoli}
\date{\today}

\begin{document}

\maketitle

\begin{abstract}

We show that the quantum computational speedup of quantum algorithms is due to their teleological character, their being evolutions toward a goal (the solution of the problem) with an attractor in the very goal they will produce in the future (the solution of the problem again).  We also show that, under the quantum cosmological assumption and for the Fine-tuned Universe version of the Anthropic Principle, the physical basis of the teleological character of quantum algorithms applies as well to the evolutions of the living for which the teleological notion was originally conceived. 
\end{abstract}

\section{Introduction}
    
  Once applied to the evolutions of the living by natural philosophy, the notion of teleological evolution---of an evolution attracted by a final goal---was dismissed with the advent of modern science for the alleged lack of a physical basis. For its finalistic character, it was also ascribed to the category of beliefs. In this work, we shall show that the relatively recent discovery of the quantum computational speedup, the fact that quantum algorithms can solve problems with fewer logical operations (computation steps) than the minimum classically possible, should change this position of science. Indeed, we shall show that the quantum speedup of quantum algorithms is due to their teleological character.

This work builds up and develops on the time-symmetric approach to the quantum computational speedup \cite{castagnoli2001,castagnoli2010,castagnoli2018,castagnoli2019,castagnoli2021,castagnoli2025}. In particular, \cite{castagnoli2019}, shows that: 

\medskip

\textit{In quantum algorithms, it is as if the problem-solver, before beginning her problem-solving action, knew in advance any of the possible halves of the information that specifies the solution of the problem she will produce and measure in the future and could use this knowledge to produce the solution with fewer computation steps; the number of computation steps needed to solve the problem in an optimal quantum way is that needed to solve it in an optimal logical way benefiting from the advanced knowledge in question}.

\medskip

We call the above \textit{the advanced knowledge rule}. In the above publications, it was given as an empirical rule holding for all the quantum algorithms examined, which comprised the major ones. In the present work, we demonstrate it. 

Let us first discuss this rule. In addition to providing a unified quantitative explanation of all quantum speedups that is absent in mainstream quantum information, it implies the teleological character of quantum algorithms. Indeed, according to it, they are evolutions toward a goal (the solution of the problem) with an attractor in the very goal they will produce and measure in the future (the solution of the problem again): Without parentheses and their contents, this can well be considered the definition of teleological evolution. This also clarifies the finalistic character of the notion. A teleological evolution is not attracted by some superior scientifically undefinable goal, as argued by the opponents of the notion, but, quite simply, by its own end product. 

Let us underline the fact that the advanced knowledge rule establishes the number of computation steps needed to solve an (oracle) problem in an optimal quantum way. Whether the quantum algorithm is optimal is another thing. We can say that the advanced knowledge rule tells us the \textit{room} for the quantum speedup.

Knowing in advance, on the part of the problem-solver, any of the possible halves of the information about the solution she will produce and measure in the future and using this knowledge to produce the solution with fewer computation steps is obviously a causal loop. As there is a plurality of ways of taking half of the information about the solution, we can say that, according to the advanced knowledge rule, the quantum speedup is explained by the mutually exclusive \textit{or} of these causal loops.

We will provide two redundant demonstrations of the advanced knowledge rule. Besides the fact that this redundancy is not to be despised due to the destabilizing character of the rule, together they provide two complementary perspectives of the role of retrocausality (more precisely, time-symmetric causality) in quantum computation.

Of course, to show that quantum algorithms are submitted to the advanced knowledge rule, we have to deal with their quantum description. First, for simplicity, we will always consider quantum algorithms for which the solution of the problem is an invertible function of the problem-setting and is determined with certainty. Second, we must observe that the current quantum description of quantum algorithms is trivially incomplete. It consists of the unitary computation of the solution, where the quantum speedup resides, and the final measurement of the latter. The canonically complete quantum description of a quantum process consists of an initial measurement, a unitary evolution, and a final measurement. 

We should first amend this incompleteness. To the quantum description of the process of solving the problem, we should add that of the upstream process of setting the problem. An initial measurement of the problem-setting in a quantum superposition (or, indifferently, mixture) of all the possible problem-settings selects one of them at random \footnote{We provide an example of what we mean for \textit{problem-setting}. Bob, the problem setter, hides a ball in a chest of $N$ drawers; Alice, the problem-solver should locate it by opening drawers. The number of the drawer with the ball selected by Bob is the problem-setting.}. The unitary computation and final measurement of the corresponding solution follow. 

We call this \textit{the ordinary quantum description of the quantum algorithm}. It is \textit{ordinary} because it is still the customary \textit{causal} quantum description of mainstream quantum mechanics, with causality always going in the same time direction. Further below, we shall see that also this quantum description is incomplete, this time in a fundamental way.

The first demonstration of the advanced knowledge rule starts with the observation that the ordinary quantum description of quantum algorithms only involves the measurement of (two) commuting observables (the problem-setting and the solution of the problem). It is thus subject to classical logic. As we shall see in the next section, the classical logic description of what the quantum algorithm does, i.e. producing the solution with fewer computation steps than possible in the classical case, is itself the advanced knowledge rule. 

Note that, with this, we would have a classical logic description of quantum algorithms that describes the advanced knowledge rule, whereas their ordinary quantum description does not. Consistently with the present picture, this should imply that the latter description is still incomplete, this time in a fundamental way. We shall see that (i) it is the same incompleteness of the quantum description lamented by Einstein et al \cite{einstein1935} in the case of quantum nonlocality and (ii) either quantum description is completed by time-symmetrizing it. In the case of quantum algorithms, this generates the quantum superposition of the causal loops whose mutually exclusive \textit{or} is implied in the advanced knowledge rule; thus, it generates the physical (quantum) correspondent of the advanced knowledge rule. In the case of quantum nolocality, the causal loops generated by the time-symmetrization exactly play the role of the hidden variables envisaged by Einstein et al.

Inspired by time-symmetric quantum mechanics \cite{aharonov1964,aharonov2007,aharonov2015}, the time-symmetrization procedure we are dealing with will be described in the next section.

The second demonstration of the advanced knowledge rule simply consists in the fact that this time-symmetrization of the ordinary quantum description of the quantum algorithm, which produces the physical correspondent of the advanced knowledge rule, turns out to be mandatory to satisfy an atemporal symmetry implicit in this quantum description---see the next section.

With this, we can go to the title of the present work. We have seen that a demonstration of the advanced knowledge rule is also a demonstration of the teleological character of quantum algorithms, indeed their being evolutions toward a goal (the solution of the problem) with an attractor in the very goal they will produce in the future (the solution of the problem again). This should be enough to dismantle the long standing belief that the notion of teleological evolution has no physical basis. Of course, it is a physical basis in the quantum world, and the teleological notion was originally applied to the macroscopic evolutions of life customarily placed in the classical world.

However, we shall show that, under the quantum cosmological assumption \cite{bojowald2011,barrow1988} and for the Fine-tuned Universe version of the Anthropic Principle \cite{barrow1988}, the physical basis of the teleological character of quantum algorithms identically applies to the evolutions of the living. 

The quantum cosmological assumption is the assumption that the universe evolves in a unitary quantum way. That version of the Anthropic principles states that the least change in the value (for values) that the fundamental physical constants have in the universe we are in would have produced a relatively trivial universe unable to develop life.

This can be formalized as follows. One should introduce two commuting observables, the value (for values) of the fundamental physical constants and the ``value of life'' (assuming, in particular, the values ``existence of sentient life'' and ``absence of life''). At the beginning of time, there would be a quantum superposition of all the possible universes, all with the same fundamental laws but for the value of the fundamental constants specific for each universe. Of course, no matter the value of the fundamental constants, the value of life would be ``absence of life'': the two observables would be unrelated. Under these fundamental laws, this quantum superposition would unitarily evolve until producing, in relatively recent times, a quantum superposition of universes, one of them---the universe we are in---containing sentient observers, the others the absence of life. In the universe we are in, the sentient observers would measure the ``value of life'', thus also selecting the value of the fundamental constants maximally entangled with it; since the latter value never changes during the evolution of any universe, all is as if it were selected at the beginning of time. 

Note that this would exactly map on the celebrated Paul Davies' cosmic bio-friendliness conjecture, according to which there would be mutual causation between the selection of the value of the fundamental constants at the beginning of time and the emergence of sentient life in relatively recent times \cite{davies2006}.

Summing up, this evolution of the quantum superposition of all the possible universes would unitarily build up (say) one-to-one correlation between the two initially unrelated observables we are dealing with, of course never changing the the value of the fundamental constants selected by the initial measurement. 

Note that this is what quantum algorithms (keeping in mind the assumption that the solution is an invertible function of the problem-setting and is determined with certainty) do. Indeed, they unitarily build up one-to-one correlation between the two (commuting) initially unrelated observables ``value of the solution'' and ``value of the problem-setting'', without ever changing the value of the problem-setting selected by the initial measurement. We shall see that this kind of building up of (quantum) correlation can always occur with quantum speedup (in the sense that there is always the room for its occurring with quantum speedup), and therefore be teleological in character. We shall also see that a teleological character of the evolution of the universe toward the value of life would be inherited by the evolution of life on Earth and, with it, those of species.

In summary, the time-symmetrization of the ordinary quantum description of quantum processes we are dealing with (and its corresponding completion) turns out to have a special unification power. Besides providing a unified quantitative explanation of all quantum speedups (the one provided by the advanced knowledge rule), it unifies the explanation of the quantum speedups with that of quantum nonlocality; eventually, under the quantum cosmological assumption, it unifies the teleological character of quantum algorithms with that of the (macroscopic) evolutions of the living for which the teleological notion was originally conceived.

The main result of the present work is, of course, its showing that quantum algorithms are teleological evolutions. As we shall see, this, in hindsight, should be most evident. If it went unnoticed until now, 41 years after the discovery of the quantum computational speedup, it is likely because the de facto non-retrocausality precept prevented to look in the needed direction. 

This result should be sufficient to dismantle the long standing notion that teleological evolutions have no physical basis. Indeed, they have it, although in the quantum world. The fact that, under the quantum cosmological assumption, this physical basis also applies to the macroscopic evolutions of the living is, of course, at least as hypothetical as the quantum cosmological assumption. Of course, this hypothetical character should not detract from the fact that quantum algorithms are teleological evolutions. This latter fact, also alone, should be enough to shake the present position of science on the teleological notion.

\section{Time-symmetric explanation of the quantum speedup}

We provide a description of quantum algorithms open to the time-symmetric coexistence of causality and retrocausality on unitary evolutions. The section is divided into subsections corresponding to the various arguments.

\subsection{Critique of the current way of viewing quantum computation}

Let us first acknowledge the fact that that of the quantum computational speedup has been one of the fundamental discoveries of the past century. Together with that of quantum communication, it has created an entirely new branch of physics. However, for the sake of clarity, we will be completely frank in criticizing what we think are the shortcomings of the current way of viewing quantum computation, in our opinion all related to an undue rejection of the notion of retrocausality.

We repeat here for convenience the advanced knowledge rule; of course, it plays an essential role in the present work.

\medskip

\textit{In quantum algorithms, it is as if the problem-solver always knew in advance, before beginning her problem-solving action, any of the possible halves of the information that specifies the solution of the problem she will produce and measure in the future and could use this knowledge to produce the solution with fewer computation steps. The number of computation steps needed in the optimal quantum case is that needed in the optimal logical case benefiting from the advanced knowledge in question.}

\medskip

This rule would offer a unified quantitative explanation of the quantum speedups that is still lacking in mainstream quantum information. To the best of our knowledge, this absence has never been acknowledged by the scientific community.

Of course, it can be said that each quantum algorithm explains its quantum speedup. However, it would be an explanation completely different for different quantum algorithms. It would be like having different explanations for different throws of a stone without knowing that they all describe a parabola. Naturally, the unification of explanations is a fundamental objective of science.

However, people currently say that there is a unified explanation of the quantum speedup; it would be that of \textit{quantum parallel computation}, the fact that many computations are performed at the same time in quantum superposition. Indeed, this captures the essential character of quantum computation, but, being a qualitative explanation of the quantum speedup, which is a quantitative feature, it cannot be considered its explanation. Of course, science essentially differs from opinion because of its quantitative character; then, settling for a qualitative explanation of a quantitative feature would seem to be a shortcoming. This is our opinion, of course. However, as a matter of fact, outside the present retrocasusal approach to the quantum speedups, in the literature, there is no trace whatsoever of any search for a unifying-quantitative explanation of them. 

Of course, the advanced knowledge rule would exactly be that unifying-quantitative explanation. Naturally, it would also be the only possible explanation, and one necessarily involving a time symmetric role of causality in unitary evolutions. The de facto non-retrocausality precept might therefore be the reason for which our search for a unified quantitative explanation of the quantum speedup never became mainstream. In this section, we shall provide the two demonstrations of the advanced knowledge rule, beginning with the one relying on the classical logic description of the quantum speedup.

\subsection{Choice of the quantum algorithm}

We have to choose a quantum algorithm to work with. We need a quantum algorithm that certainly gives a quantum speedup with respect to what is classically possible. Now, most quantum algorithms yield a (gigantic) quantum speedup with respect to the best of their known classical counterparts, but it is not known whether the latter is the best classically possible. Excluding elementary one-computation-step quantum algorithms whose dimensions do not scale up, we find our choice limited to Grover's quantum search algorithm \cite{grover1996}. It solves the following problem: Bob, the problem-setter, hides a ball in a chest of $N$ drawers. Alice, the problem-solver is to locate it by opening drawers. The number of drawer openings needed by the (optimal) Grover algorithm is on the order of $\sqrt{N}\label{eq:square-root-of-N}$ against the order of $N$ of the best classical algorithm, and hence its \textit{quadratic} quantum speedup.

Besides yielding a well determined quantum speedup with respect to what is classically possible, Grover algorithm has the following convenient features: (i) In Long's version \cite{long2001}, it always provides the solution with certainty; this simplifies things. (ii) In it, the solution is an invertible function of the problem-setting; this simplifies its time-symmetrization. (iii) It is optimal in character; this, of course, constitutes an important reference. (iv) It is a most universal quantum algorithm: it can easily be adapted to solve any NP problem with a quadratic quantum speedup; more than a single quantum algorithm, it is an entire test bed on which to test the present theory of the quantum speedup.

In the following, we will always work with the simplest, four drawer instance of Grover algorithm. Everything will immediately generalize to any number of drawers. In \cite{castagnoli2019}, we have seen that it also generalizes to all the major quantum algorithms.

Let us recall the quantum description of quantum algorithms extended to the process of solving the problem already provided in the \textit{Introduction}:

\medskip

\textit{An initial measurement in a quantum superposition (or, indifferently, mixture) of all  the possible problem-settings selects one of them at random. The unitary computation and the final measurement of the corresponding solution follow. }

\medskip

We provide its formalization.
\medskip

We number the four drawers in binary notation: $00 ,01 ,10 ,11$. We will need two quantum registers of two \textit{quantum bits} each:

\medskip

A register $B$ under the control of the problem-setter Bob contains the problem-setting observable $\hat{B}$ (the number of the drawer in which the ball is hidden) of eigenstates and corresponding eigenvalues:

\medskip \medskip

$\vert 00 \rangle _{B}$ and $00$; $\vert 01 \rangle _{B}$ and $01$; $\vert 10 \rangle _{B}$ and $10$; $\vert 11 \rangle _{B}$ and $11$. 

\medskip 

Measuring $\hat{B}$ in a quantum superposition (or, indifferently, mixture) of all the possible problem-settings selects one of them at random.

\medskip
              
A register $A$ under the control of the problem-solver Alice contains the number of the drawer opened by her. The content of register $A$ is the observable $\hat{A}$  of eigenstates and corresponding eigenvalues:

\medskip

$\vert 00 \rangle _{A}$ and $00$; $\vert 01 \rangle _{A}$ and $01$; $\vert 10 \rangle _{A}$ and $10$; $\vert 11 \rangle _{A}$ and $11$.

\medskip

Measuring
$\hat{A}$  at the end of the quantum algorithm yields the number of the drawer with the ball selected by the initial measurement, namely the solution of the problem. Note that
$\hat{B}$  and
$\hat{A}$  commute.

\medskip

The quantum description of the extended quantum algorithm is: 

\medskip

\begin{equation}\begin{array}{ccc}\begin{array}{c}\;\text{time }t_{1}\text{, meas. of}\;\hat{B}\end{array} & t_{1} \rightarrow t_{2} & \text{time }t_{2}\text{, meas. of}\;\hat{A} \\
\, & \, & \, \\
\begin{array}{c}\left (\vert 00 \rangle _{B} +\vert  01 \rangle _{B} +\vert 10 \rangle _{B} +\vert  11 \rangle _{B}\right ) \\
\left (\vert 00 \rangle _{A} +\vert 01 \rangle _{A} +\vert 10 \rangle _{A} +\vert 11 \rangle _{A}\right )\end{array} & \, & \, \\
\Downarrow  & \, & \, \\
\vert 01 \rangle _{B}\left (\vert 00 \rangle _{A} +\vert 01 \rangle _{A} +\vert 10 \rangle _{A} +\vert 11 \rangle _{A}\right ) &  \Rightarrow \hat{\mathbf{U}}_{1 ,2} \Rightarrow  & \vert 01 \rangle _{B}\vert  01 \rangle _{A}\end{array}\tag{Table I}
\end{equation}

\medskip

Here and in the following, we disregard normalization.

We should start from the quantum state in the upper-left corner of Table I diagram and then follow the vertical and horizontal arrows. Initially, at time $t_{1}$ (upper-left corner of the diagram), both registers, each by itself, are in a quantum superposition of all their basis vectors. By measuring
$\hat{B}$  (the content of $B$) in this initial state, Bob selects the number of the drawer with the ball at random; say it comes out $01$: see the state of register $B$ under the vertical arrow\protect\footnote{By the way, Bob would be free to unitarily change at will the sorted out number of the drawer with the ball. We omit this possible operation that would change nothing here.}.

There is then the unitary computation of the solution, by the unitary transformation $\hat{\mathbf{U}}_{1 ,2}$. Of course, the latter produces the solution of the problem never changing the problem-setting. 

Note that we will never need to know $\hat{\mathbf{U}}_{1 ,2}$ (i.e. Grover algorithm). We will only need to know what it does. Between the outcome of the initial measurement and that of the final measurement, it unitarily builds up one-to-one correlation between the two initially unrelated observables $\hat{B}$ and $\hat{A}$ without ever changing the eigenvalue of the observable selected by the initial measurement.

At the end of Alice's problem-solving action, at time $t_{2}$, register $A$ contains the solution of the problem, i.e. the number of the drawer with the ball---bottom-right corner of the diagram. 

Eventually, by measuring
$\hat{A}$  (the content of register $A$), which is already in one of its eigenstates, Alice acquires the solution with probability $1$---the quantum state in the bottom-right corner of the diagram naturally remains unaltered throughout this measurement.

\subsection{Relativizing the extended quantum description of the quantum algorithm with respect to the problem-solver}

The quantum description of the quantum algorithm of Table I is with respect to the customary external observer, it cannot be that with respect to the problem-solver Alice. It would tell her the solution of the problem before she begins her problem-solving action---see the content of register $B$ in the state under the vertical arrow. Now, we will need to work with the quantum description of the quantum algorithm with respect to her: naturally, Alice's advanced knowledge of one of the possible halves of the information about the solution must appear in the (completed) quantum description of the quantum algorithm with respect to her. Since Alice's problem-solving action never changes the number of the drawer with the ball selected by the initial measurement and the observables $\hat{B}$ and $\hat{A}$  commute, describing this concealment is simple. It suffices to postpone after the end of Alice's problem-solving action the projection of the quantum state associated with the initial measurement of $\hat{B}$  (or, indifferently, to postpone the very measurement of
$\hat{B}$). The result is the following quantum description of the quantum algorithm with respect to Alice:

\medskip

\begin{equation}\begin{array}{ccc}\begin{array}{c}\text{time}\text{ }t_{1}\text{, meas. of}\;\hat{B}\end{array} & t_{1} \rightarrow t_{2} & \text{\text{time }\text{}}t_{2}\text{, meas. of}\;\hat{A} \\
\, & \, & \, \\
\begin{array}{c}\left (\vert 00 \rangle _{B} +\vert  01 \rangle _{B} +\vert 10 \rangle _{B} +\vert  11 \rangle _{B}\right ) \\
\left (\vert 00 \rangle _{A} +\vert 01 \rangle _{A} +\vert 10 \rangle _{A} +\vert 11 \rangle _{A}\right )\end{array} &  \Rightarrow \hat{\mathbf{U}}_{1 ,2} \Rightarrow  & \begin{array}{c}\vert 00 \rangle _{B}\vert  00 \rangle _{A} +\vert 01 \rangle _{B}\vert  01 \rangle _{A} + \\
\vert 10 \rangle _{B}\vert  10 \rangle _{A} +\vert 11 \rangle _{B}\vert  11 \rangle _{A}\end{array} \\
\, & \, & \Downarrow  \\
\, & \, & \vert 01 \rangle _{B}\vert  01 \rangle _{A}\end{array}\tag{Table II}
\end{equation}

\medskip 

 To Alice, the initial measurement of \textbf{$\hat{B}$} leaves the initial state of register $B$ unaltered. At the beginning of her problem-solving action, she is completely ignorant of the number of the drawer with the ball selected by Bob---see the state of register $B$ on the left of $ \Rightarrow \hat{\mathbf{U}}_{1 ,2} \Rightarrow $.

 The unitary transformation $\hat{\mathbf{U}}_{1 ,2}$ is then performed for a quantum superposition of the four possible numbers of the drawer with the ball. The state at the end of it---on the right of the horizontal arrows---is a quantum superposition of tensor products, each a possible number of the drawer with the ball in register $B$ tensor product the corresponding solution (that same number) in register $A$. Since
$\hat{B}$ and
$\hat{A}$  commute, the final measurement of
$\hat{A}$  must project this superposition on the tensor product of the number of the drawer with the ball initially selected by Bob and the corresponding solution---bottom-right corner of Table II diagram.

Note that this relativization relies on relational quantum mechanics \cite{rovelli1996,fuchs2016} where the quantum state is relative to the observer. In the following, we will call this quantum description of the quantum algorithm with respect to the problem-solver its \textit{ordinary quantum description}. It is ``ordinary'' because it is still the customary ``causal'' quantum description of mainstream quantum mechanics---where causality always goes in the same time-direction.

\subsection{First demonstration of the advanced knowledge rule}

We can now go to the first demonstration of the advanced knowledge rule. It starts with the observation that, since the two measurements of the quantum algorithm are of commuting observables (the problem-setting and the solution of the problem), the latter, and what it does, are subject to classical logic. The demonstration simply consists in describing in a classical logic way what the quantum algorithm does---computing the solution of the problem with a number of computation steps that are less than the minimum possible in the classical case. As we shall soon see, this very description is the advanced knowledge rule.

First, let us recall what is the quantum speedup of the simplest instance of Grover algorithm. The problem to be solved is as follows. Bob, the problem-setter, hides a ball in a chest of four drawers. Alice, the problem-solver, is to locate the ball by opening drawers. In the worst classical case, she has to open three drawers: if the ball is only in the third drawer opened, she has solved the problem; if not, it must be in the only drawer not yet opened, and she has solved the problem of locating the ball as well. Instead, Grover algorithm (its simplest instance) always solves the problem by opening just one drawer, in a quantum superposition of the four possible drawer numbers. Obviously, always solving the problem by opening just one drawer would be impossible in the classical case: indeed, we are dealing with a quantum computational speedup.

There is a classical logic equivalent of the fact that the quantum algorithm always solves the problem by opening a single drawer. By the way, since the fact that a single drawer is opened is logically true in itself, we can forget the specification that this is done for a quantum superposition of the four drawer numbers. 

The logical equivalence is as follows. Logically, it is as if the problem-solver always knew in advance, before beginning her problem-solving action, that the ball is in a given pair of drawers (in a mutually exclusive way one of the three pairs of drawers in which the ball must be) and took advantage of this knowledge to solve the problem by opening just one of them. In fact, if the ball is in the opened drawer, she has solved the problem; if not, it must be in the other drawer and she has solved the problem of locating the ball as well, in either case having opened just one drawer. By the way, this also means knowing in advance half of the information that specifies the solution of the problem that will be produced and measured in the future.

Let us open a parentheses to observe that we are not talking about the quantum algorithm as it were a classical algorithm. We are making a classical logic analysis of what the quantum algorithm does. In other words, here, classical logic should not be confused with classical physics, which might be a temptation in the present case.

Always knowing in advance that the ball is in a given pair of drawers is logically necessary to always solve the problem by opening just one drawer. Then the question becomes: from where does Alice, before beginning her problem-solving action, always get that information about the solution? Since, for her, the only possible source of it is the solution she will produce and measure in the future, the only possible answer is that the information in question retrocausally comes to her from her future measurement of the solution. This is formalized in the next subsection; however, before looking at it, one should read the description of the time-symmetrization of the quantum algorithm that we shall provide in a moment.

Note that we are dealing with causal loops. Alice knows in advance one of the possible halves of the information about the solution she will produce and measure in the future (that the ball is in two rather than four drawers) and can use this knowledge to produce the solution by opening just one drawer, against the up to three of the classical case. There is a causal loop for each way of taking half of the information about the solution. Therefore, we can say that this quantum speedup logically implies the mutually exclusive \textit{or} of the causal loops we are dealing with. Of course, this is the advanced knowledge rule for the present instance of Grover algorithm, logically implied by the sheer existence of the quantum speedup. 

By the way, note that we never needed to know how the quantum algorithm achieves the quantum speedup, we only needed to know that it achieves it.

This explanation of the quantum speedup immediately generalizes to any number $N$ of drawers, namely to the full fledged Grover algorithm. As is well known, the number of drawer openings needed by the (optimal) Grover algorithm is on the order of $\sqrt{N}\label{eq:square-root-of-N}$ against the order of $N$ of the best classical case, hence its \textit{quadratic} quantum speedup. This is logically equivalent to knowing in advance, in a mutually exclusive way, that the ball is in one of the possible tuples of $\sqrt{N}$ drawers and benefiting from this knowledge to locate the ball in the best logical way by opening the order of $\sqrt{N}$ drawers. Of course, this is what the optimal Grover algorithm does. Also in this case, using the advanced knowledge of half of the information about the solution in an optimal logical way yields the optimal quantum speedup of Grover algorithm. This is, of course, the advanced knowledge rule for the full fledged Grover algorithm.

Note that, in equivalent terms, the computational complexity of the problem to be solved reduces from searching the ball in $N$ drawers to searching it in $\sqrt{N}\label{eq:square_root_of_N_alt1}$ drawers. In still equivalent terms, the dimension of the Hilbert space in which the search for the solution takes place reduces from $N$ to $\sqrt{N}\label{eq:square_root_of_N_alt2}$.

Holding for Grover algorithm, the above retrocausal explanation of the quantum speedup must hold for the process of solving with a quadratic quantum speedup any NP problem. We shall see that there are further generalizations in the following of this section.

By the way, we have seen that the advanced knowledge rule yields us the room for achieving a quantum speedup. Interestingly, exploiting this room may not be difficult: \cite{shenvi2002} shows that, in quantum search algorithms, even a random quantum search for the solution yields a just suboptimal quantum speedup.

\subsection{Completing the ordinary quantum description of the quantum algorithm by time-symmetrizing it}

Consistently with the picture that we are developing, the above subsection also says that the ordinary quantum description of the quantum algorithm, which obviously does not describe the advanced knowledge rule, is incomplete. Of course, to bring it in line with its classical logic description, it should be completed. It turns out that this completion is done by a time-symmetrization of the ordinary quantum description. The kind of time-symmetrization adopted will be justified in two redundant ways: One is, indeed, that it brings the ordinary quantum description of the quantum algorithm in line with its classical logic description. The other is that it will be mandatory to satisfy an atemporal symmetry implicit in the ordinary quantum description of the quantum algorithm.

Let us describe this time-symmetrization. We are always in the case where the solution is determined with certainty and is an invertible function of the problem-setting. 

Naturally, in the ordinary quantum description of the quantum algorithm, the initial measurement of the problem-setting selects the information that specifies the problem-setting and the corresponding solution. To time-symmetrize this ordinary quantum description, we should evenly share the selection of this information between the initial measurement of the problem setting and the final measurement of the solution. As there is a plurality of ways of evenly sharing, we should take their quantum superposition.

In each time-symmetrization instance, the selection performed by the initial measurement should propagate forward in time by the unitary computation of the solution $\hat{\mathbf{U}}_{1 ,2}$ until becoming the state immediately before the final measurement. Time-symmetrically, the complementary selection performed by the final measurement should propagate backward in time by the inverse of the computation of the solution $\hat{\mathbf{U}}_{1 ,2}^{\dag }$ until replacing the previous outcome of the initial measurement. This backward in time propagation, which inherits both selections, is an instance of the time-symmetrized quantum description. 

This is formalized as follows: 

We consider the time-symmetrization instance where: 

\medskip

(i) The initial measurement of $\hat{B}$  reduces to the measurement of
$\hat{B}_{l}$, the left bit of the two-bit number contained in register $B$; say that it randomly selects left bit $0$. 

\medskip

(ii) The final measurement of
$\hat{A}$  should correspondingly reduce to that of $\hat{A_{r}}$, the right bit of the two-bit number contained in register $A$; say that it randomly selects right bit $1$. The randomly selected number of the drawer with the ball is thus $01$. 

\medskip

In this particular time-symmetrization instance, the ordinary quantum description of the quantum algorithm of Table II becomes:

\medskip 

\begin{equation}\begin{array}{ccc}\begin{array}{c}\;\text{time }t_{1}\text{, meas. of}\;\hat{B}_{l}\end{array} & t_{1 \rightleftarrows }t_{2} & \text{time }t_{2}\text{,meas. of}\;\hat{A_{r}} \\
\, & \, & \, \\
\begin{array}{c}\left (\vert 00 \rangle _{B} +\vert  01 \rangle _{B} +\vert 10 \rangle _{B} +\vert  11 \rangle _{B}\right ) \\
\left (\vert 00 \rangle _{A} +\vert 01 \rangle _{A} +\vert 10 \rangle _{A} +\vert 11 \rangle _{A}\right )  \end{array} &  \Rightarrow \hat{\mathbf{U}}_{1 ,2} \Rightarrow  & \begin{array}{c}\vert 00 \rangle _{B}\vert  00 \rangle _{A} +\vert 01 \rangle _{B}\vert  01 \rangle _{A} + \\
\vert 10 \rangle _{B}\vert  10 \rangle _{A} +\vert 11 \rangle _{B}\vert  11 \rangle _{A}\end{array} \\
\, & \, & \Downarrow  \\
\begin{array}{c}\left (\vert 01 \rangle _{B} +\vert  11 \rangle _{B}\right ) \\
\left (\vert 00 \rangle _{A} +\vert 01 \rangle _{A} +\vert 10 \rangle _{A} +\vert 11 \rangle _{A}\right )\end{array} &  \Leftarrow \hat{\mathbf{U}}_{1 ,2}^{\dag } \Leftarrow  & \vert 01 \rangle _{B}\vert  01 \rangle _{A} +\vert 11 \rangle _{B}\vert  11 \rangle _{A}\end{array}\tag{Table III}
\end{equation}

\medskip

The projection of the initial state of register $B$ due to the measurement of
$\hat{B}_{l}$---top-left corner of the diagram---is postponed after Alice's problem-solving action, namely outside the present table which is limited to that action. With this postponement, the initial state of register $B$ goes unaltered through the initial measurement of
$\hat{B}_{l}$  becoming, by $\hat{\mathbf{U}}_{1 ,2}$, the quantum superposition of tensor products above the vertical arrow. Alice's final measurement of
$\hat{A_{r}}$, which (in this time-symmetrization instance) selects the value $1$ of the right bit of register $A$, projects the state above the vertical arrow on the state below it. Then (in causal order), this final measurement  outcome propagates backward in time by $\hat{\mathbf{U}}_{1 ,2}^{\dag }$ until becoming the definitive outcome of the initial measurement. This backward in time propagation---the bottom line of the diagram---is an instance of the time-symmetrized quantum algorithm.

Note that, for the mathematical equivalence between $ \Leftarrow \hat{\mathbf{U}}_{1 ,2}^{\dag } \Leftarrow $ and $ \Rightarrow \hat{\mathbf{U}}_{1 ,2} \Rightarrow $, this  bottom line can also be read in the usual left to right way:

\medskip

\begin{equation}\begin{array}{ccc}\text{time }t_{1} & t_{1} \rightarrow t_{2} & \text{time }t_{2} \\
\, & \, & \, \\
\begin{array}{c}(\vert 01 \rangle _{B} +\vert  11 \rangle _{B}) \\
\left (\vert 00 \rangle _{A} +\vert 01 \rangle _{A} +\vert 10 \rangle _{A} +\vert 11 \rangle _{A}\right )  \end{array} &  \Rightarrow \hat{\mathbf{U}}_{1 ,2} \Rightarrow  & \vert 01 \rangle _{B} \vert 01 \rangle _{A} +\vert  11 \rangle _{B}\vert 11 \rangle _{A}\end{array}\tag{Table IV}
\end{equation}

\medskip

Note that, in each time-symmetrization instance (bottom line of the diagram of Table III or the diagram of Table IV), it is as if the problem-solver knew in advance one of the possible halves of the information about the solution she will produce and measure in the future (that the right bit of the number of the drawer with the ball is 1): see the state of register $B$ on the left of the bottom line of Table III diagram, or on the left of Table IV diagram. She can then use this knowledge to compute the solution with fewer computation steps: this is still described by the bottom line of Table III diagram, read from left to right, or the diagram of Table IV.

We can also see that, in each time-symmetrization instance, the computational complexity of the problem to be solved by Alice quadratically reduces from the problem of locating the ball in one of $4$ drawers to that of locating it in $2 =\sqrt{4}$ drawers. In still equivalent terms, the dimension of the Hilbert space in which the search for the solution takes place quadratically reduces from $4$ basis vectors to $2 =\sqrt{4}$ basis vectors.

Eventually, the time-symmetrized quantum description is the quantum superposition of all the time-symmetrization instances.

In summary, the causal loop generated by each instance of the time-symmetrization process is the zigzag diagram of Table III. We can see that it is one of the causal loops whose mutually exclusive $or$ is logically implied by the sheer existence of the quantum speedup. The state of register $B$  in the top-left part of the diagram tells us that Alice, before beginning her problem-solving action, is completely ignorant of the number of the drawer with the ball. Through the zig-zag, this state changes into the state of register $B$ in the bottom-left part of the diagram, which tells us that Alice, before beginning her problem-solving action, knows in advance that the ball is in drawer either $01$ or $11$. She can then solve the problem in an optimal logical way by opening just one of the two drawers, indeed as in the classical logic description of what the quantum algorithms does---see the bottom line of Table III diagram read from left to right or the diagram of table IV.

Eventually, we should take the quantum superposition of all the time-symmetrization instances. The time-symmetrized quantum description of the quantum algorithm (i.e. its completed quantum description) becomes the quantum superposition of the causal loops whose mutually exclusive \textit{or} is logically implied by the quantum speedup. Thus completed, also the quantum description of the quantum algorithm describes the advanced knowledge rule (more precisely, it describes its physical correspondent).

\subsection{Second demonstration of the advanced knowledge rule}

The second demonstration of the advanced knowledge rule simply comes from the observation that the time-symmetrization of the ordinary quantum description of the quantum algorithm, which provides the (physical correspondent of the) advanced knowledge rule, is mandatory also for another reason, as follows.

In the ordinary quantum description of the quantum algorithm, of course, the unitary computation of the solution never changes the corresponding problem-setting. Therefore, one should be free to postpone the initial measurement of the problem-setting along the subsequent unitary computation of the solution until it becomes simultaneous with the final measurement of the solution. At this point, we find ourselves with the simultaneous measurements of two maximally entangled observables (the problem-setting and the solution, one an invertible function of the other), and the time-symmetrization we are dealing with must already be in place to represent the perfect (atemporal) symmetry between these two measurements. Of course, there would be no reason why a measurement selected more information than the other. 

As we have just seen, this time symmetrization generates the quantum superposition of the causal loops whose mutually exclusive \textit{or} is logically implied by the sheer existence of the quantum speedup---the physical correspondent of the advanced knowledge rule. This is indeed a second demonstration of the advanced knowledge rule. Note that it also justifies the form of time-symmetrization adopted, of course, in both the present and the previous subsection.

By the way, this second demonstration is reminiscent of Wheeler and Feynman's \cite{wheeler1949} time-symmetric absorber theory; we would say: with the advantage that the time-symmetrization we are presently dealing with relies on the reversibility of unitary quantum evolutions, which, unlike classical reversibility, is such in both an ideal and objective way.

\subsection{Origin of the mutual causality}

A quantum algorithm is a relatively complex process. A natural question is whether the mutual causality implicit in the classical logic description of its overall result, the quantum speedup, derives from some more elementary part of it. Of course, in its ordinary quantum description, the unitary computation of the solution is an alternation of generations of quantum state superpositions and quantum interferences. Since the latter have a classical analog, the essentially quantum generation of a quantum state superposition is the suspect.

Let us first provide a very intuitive exposition of the fact that the classical logic description of a state superposition necessarily involves the coexistence of causality and retrocausality. We resort to a prescientific notion of state superposition. It is the story of the monk seen by his disciples on the two banks of the river at the same time. We should provide a classical logic description of this. Being on one and the other bank is possible in two different times but, at the same time, from the ordinary classical logic standpoint, they should be two mutually exclusive possibilities. Until now, being on the two banks at the same time would seem to be a logical impossibility. However, let us note that it would be impossible in a classical logic description closed to the coexistence of causality and retrocausality. Opening classical logic to this coexistence completely changes the situation\footnote{Classical logic is neutral for what concerns the notion of causality. Associating it with a causality that always goes in the same time-direction, as customary, must be consequent to the de facto non-retrocausality precept. Being neutral, classical logic should as well go along with the coexistence of causality and retrocausality.}. Indeed, the monk could have gone to one of the two banks along a certain path; then (\textit{then} in causal order) he could have gone backward in time undoing the path just followed (like in a reversed film); then, going forward in time again, he could have reached the other bank along a different path. This causality zigzag, of course, makes his being on the two banks at the same time logically possible.

Let us port the above to the notion of quantum state superposition. Think of a photon going in quantum superposition through the two branches of a Mach-Zehnder interferometer. The two branches should be tuned in such a way that the photon always ends up in the same detector, so that we have a (of course, time-reversible) unitary evolution that goes from before the photon enters the interferometer to immediately before its detection. 

The classical logic description of this quantum superposition can be as follows. In entering the interferometer, the photon takes, say, the upper branch until it reaches the state immediately before detection (by the detector in question); of course, this also provides the phase of the photon along the path followed. Then (in causal order), the photon goes backward in time undoing the path just followed until before entering the interferometer. Then, it goes forward in time again taking the lower branch until reaching the detector in question, and this time, be detected. The zigzag path followed in this way by the photon through the interferometer, with the associated photon phases, of course, describes in a classical logic way the fact that the photon goes through the interferometer in a quantum superposition of the upper and lower branches, eventually ending up in that detector. It should be clear that the classical logic description of the generation of a quantum state superposition necessarily involves the coexistence of the two time directions of causality along the same unitary quantum process.

We should think that the classical logic description of the sequence of generations of quantum state superpositions and interferences that constitutes the unitary computation of the solution simplifies into the advanced knowledge rule. 

Summarizing, at the root of the mutual causality we are dealing with there would be the classical logic description of the fundamental notion of generation of a quantum state superposition. By the way, note that this is not an idle speculation as it might even seem. Indeed, we are dealing with a form of coexistence of the two time-directions of causality with huge consequences, as that of providing a physical basis to the notion of teleological evolution.

\subsection{Generalization} 

In the second demonstration of the advanced knowledge rule, we have seen that the ordinary quantum description of a quantum algorithm whose solution is an invertible function of the problem-setting and is determined with certainty is incomplete and must be completed by time symmetrizing it. Let us note that the necessity of this time-symmetrization does not depend on the fact that we are dealing with a quantum algorithm; there is this immediate generalization:

\medskip

\textit{The ordinary quantum  description of any unitary evolution between two one-to-one correlated measurement outcomes that does not change the eigenvalue selected by the initial measurement is incomplete and is completed by time-symmetrizing it.}

\medskip

Indeed, as one can see, to this kind of quantum process one can apply the argument for which its ordinary quantum description must be time-symmetrized to satisfy the atemporal symmetry implicit in it. Let us call this rule \textit{the quantum correlation time-symmetrization rule}.

In the case that the unitary evolution builds up one-to-one correlation between two initially unrelated observables, as in quantum search algorithms, this building up can always occur with quadratic quantum speedup, in the sense that there is always the room for its occurring with this quantum speedup. This is what the time-symmetrization of its ordinary quantum description says---see Subsection 2.5. Of course, in the case where we are not dealing with a quantum algorithm, the notion of advanced knowledge of one of the possible halves of the information about the solution must be replaced by the equivalent notion of quadratic reduction of the dimension of the Hilbert space in which the search for that one-to-one correlation takes place, from $N$ to $\sqrt{N}\label{eq:square_root_of_N}$.

The case in which the one-to-one quantum correlation is not built up by the unitary evolution because it preexists it (more precisely, what should preexist is the maximum entanglement between the observables of the two future measurements) is, for example, that of quantum nonlocality. Here, the unitary evolution spatially separates the subsystems of two already maximally entangled observables. With the two measurement apparatuses aligned and the spatial separation process preceding the first measurement, this responds to the quantum correlation time-symmetrization rule. 

As shown in \cite{castagnoli2021}, the time-symmetrization of the ordinary quantum description of this quantum nonlocality yields a quantum superposition of causal loops that locally explain it and play the role of the hidden variables envisaged by Einstein et al \cite{einstein1935}. Indeed, these causal loops locally go from one to the other measurement outcome via the time the two subsystems were not yet space separate, where they locally tell the subsystem of the not yet measured observable the state to assume in view of the future measurement. Note that this is exactly what Einstein et al required from their hidden variables. That was, of course, without resorting to any mutual-causality, but under a deterministic view of the measurement process equivalent in this case. Of course, this should also mean that the incompleteness we are presently ascribing to the ordinary quantum description of quantum algorithms is the same that Einstein et al ascribed to the quantum description of quantum nonlocality.

By the way, also the fact that the present explanation of the quantum speedup demonstrates the incompleteness of the quantum description highlighted by Einstein et al might be destabilizing. Indeed, after highlighting for the first time quantum nonlocality and inspiring fundamental works like Bell's theorem \cite{bell1964} and Ekert's quantum cryptography based on it \cite{ekert1991}, their argument had been dismissed by the physicists' community because of the alleged absence of a satisfactory way of completing the quantum description by means of local hidden variables. In particular, the argument of the incompleteness of the quantum description has been forgotten.

\subsection{The possible teleological character of the evolutions of the living}

Let us show how the teleological character of quantum algorithms could be ported to the (macroscopic) evolutions of the living for which the teleological notion was originally conceived. 

The first observation must be about the enormous difference between quantum algorithms and the evolutions of life. Of course, quantum algorithms belong to the microscopic world of quantum mechanics; as things are now, they reach a size of a thousand quantum bits and a duration of a few dozens of microseconds; more than that, quantum decoherence prevails. The evolutions of life, instead, can have the size of the earth and have duration of billions of years. However, this distance is enormous only under the customary classical vision of macroscopic reality. Under the quantum cosmological assumption that the universe evolves in a unitary quantum way \cite{bojowald2011,barrow1988}, it becomes irrelevant. We should keep in mind the literally perfect time-reversibility of unitary quantum evolutions.

In the \textit{Generalization} subsection, we have seen that any unitary quantum evolution that builds up one-to-one correlation between two initially unrelated observables without ever changing the eigenvalue of the observable selected by the initial measurement can always occur with quadratic quantum speedup, in the sense that there is always the room for its occurring with quantum speedup.

Let us show that, under the quantum cosmological assumption and for the Fine-tuned Universe version of the Anthropic Principle \cite{barrow1988}, the unitary quantum evolution of the universe toward the value of life is of this kind.

That version of the Anthropic Principle is about the extraordinary theoretical fact that the least change of the value (for values) that the fundamental physical constants have in the universe we are in would have produced a relatively trivial universe unable to develop life. Under the quantum cosmological assumption, this can be represented as follows. 

We should introduce two commuting observables of the same size: the value of the fundamental constants and the ``value of life''. The latter should assume, in particular, the values ``absence of life'' and ``existence of sentient life''. 

We should then consider a quantum superposition (or, indifferently, mixture) of all the possible universes, each with its own value of the fundamental constants, all obeying the same fundamental laws (but for a specific value of the fundamental constants). Of course, at the beginning of time, these two observables would be unrelated: no matter the value of the fundamental constants, the value of life must be ``absence of life''. 

We should now consider the unitary evolution of this quantum superposition of all possible universes toward the value of life. At the time of today, this evolution must have built up maximal entanglement (say, one-to-one correlation) between the value of the fundamental constants since the beginning of time and the value of life, of course, without ever changing the former value. By the way, according to cosmology, at the time of today, the absence of life in the universes that cannot produce it would already be definitive. 

As we have seen in the \textit{Generalization} subsection, this kind of building up of quantum correlation can always occur with quadratic quantum speedup (in the sense that there is always the room for it), and therefore be teleological in character. 

In the universe we are in, the evolution of the universe toward the value of life becomes evolution of the universe toward life; of course, at the time of today there are the sentient observers who measure the value of life, also selecting the value of the fundamental constants maximally entangled with it. By the way, since the latter value is never changed by the unitary evolution of the universe toward life, all is as if it were measured at the beginning of time. We could also think that, if the search for the value of life by the evolution of the universe were random, it would have actually occurred with quantum speedup.

Summarizing, the evolution of the universe toward life could occur with quantum speedup and consequently be teleological in character. Let us show that this possibility would then be inherited by the evolution of life on Earth and the evolution of species. 

In principle, the quantum description of the evolution of life on Earth can be derived from that of the evolution of the universe toward life by tracing out all the rest. Conversely, we can speak of an image of the evolution of life on Earth in the evolution of the universe toward life. Then, first, evolution times must be the same for the evolution of life on Earth and its image in the evolution of the universe. Second, the amount of physical resources used by the evolution of life on Earth cannot exceed the amount of physical resources used by its image in the evolution of the universe. In fact, the latter resources comprise the former. Summing up, under the quantum cosmological assumption, the evolution of life on Earth could occur with quantum speedup (in the sense that there would always be the room for it) and consequently be teleological in character; it might actually occur with quantum speedup if the search for life by the evolution of the universe toward life were random. Then, just for consistency, the evolutions of species would also be the same.

Not to hide behind a finger, let us exemplify what a teleological character of the evolution of life would imply; of course, it would be very destabilizing with respect to the present way of viewing both the evolution of life and physical reality.

It is known that, for many millions of years, the ancestors of the birds were feathered dinosaurs that did not fly. Very light, strong but flexible, cave and air-proof, from the fossils found, even at that time feathers were ideal for flight, very close to those of the future birds. So, why anticipating something designed for flight many millions of years before it? 

The usual explanation, of course, is causal. First, an ancestor of feathers appeared by random DNA mutation, then feathers evolved as a good thermal insulator and perhaps to speed escape from predators; only eventually (by sheer coincidence) they turned out to be what was needed to conquer the air.

All this (without the parentheses and its content) could remain true in the teleological view with just one, but very significant, addition: even at the time of the non-flying dinosaurs, those random mutations that would have permitted in the future to the conquest of the air would have more likely occurred. In our possibly biased opinion, this mutual-causality explanation of the anticipation in question would be much more convincing than the customary causal explanation. Indeed, the feathers of the fossils of those non-flying dinosaurs are almost identical to those of their flying descendants. Our preprint \cite{castagnoli2025} provides a more comprehensive discussion of the implications of the teleological notion.

Another consequence of the present theory is that, according to it, a teleological evolution is an evolution endowed with quantum speedup. Of course, this is a feature of teleological evolutions completely absent in their historical notion. It could contribute to explaining why some steps of the evolutions of the living have been so rapid to defy Darwinian gradualism, see \cite{laland2014}. As is well known, have certainly been of this kind, in particular, the Cambrian Explosion and modern ecological adaptation.

Of course, being under the quantum cosmological assumption, all the above is hypothetical. This is to decouple it from the teleological character of quantum algorithms, which we would consider watertight.

\section{Conclusions}

The present work goes against a long-standing non-retrocausality tradition of physics. More specifically, since causality could well go backward in time along a unitary quantum evolution that is both objectively and ideally time-reversible, we should say that, in the quantum world, this tradition amounts to asserting the mutual exclusivity between the two possible time directions of causality along unitary evolutions.

We would observe that just opening up to the coexistence of these two possible time directions of causality makes the present notion of teleological evolution obvious. Always locating a ball hidden in one of four drawers by opening a single drawer, as the quantum algorithm does, is obviously equivalent to knowing in advance that the ball is, in a mutually exclusive way, in one of the three pairs of drawers in which it must be, and benefiting from this advanced knowledge about the solution that will be produced in the future to produce the solution by opening a single drawer. Most simply, this logical equivalence tells us that the process of producing the solution of the problem is attracted from the solution it will produce in the future, it is indeed teleological in character. We have seen how this immediately generalizes to any number of drawers and most general forms of quantum correlation. 

Eventually, under the quantum cosmological assumption and for the Fine-tuned Universe version of the Anthropic Principle, the teleological character of quantum algorithms can be ported to the macroscopic evolutions of the living for which the teleological notion was originally conceived.

We would believe that, despite its strongly destabilizing character, for its evidence in hindsight, the present scientific notion of teleological evolution deserves to be discussed.

\section*{Acknowledgements}

Thanks are due to Eliahu Cohen, Artur Ekert, Avshalom Elitzur and David Finkelstein for useful comments, Daniel Sheehan for organizing the San Diego AAAS Pacific Division conferences on retrocausality, a far-looking forum for the discussion of frontier, also unorthodox, physics, and Mario Rasetti for organizing the first series of international workshops on quantum communication and computation, the 1991--96 Elsag Bailey--ISI Turin Villa Gualino workshops.

\end{document}